# Projective Decomposition and Matrix Equivalence up to Scale

Concepts for a New Data Normalization Methodology

Max Robinson
Institute for Systems Biology
401 Terry Ave
Seattle, WA 98109

## Abstract

A numerical data matrix may be seen simply as a means of organizing observations of a system into rows in one manner (*e.g.,* by measured "object"), and into columns by another (*e.g.,* by measured "variable") so that the observations can be analyzed with mathematical tools. As a mathematical object, however, a matrix defines a linear mapping between points representing weighted combinations of its rows (the row vector space) and points representing weighted combinations of its columns (the column vector space). From this perspective, a data matrix defines a relationship between the information that labels its rows and the information that labels its columns, and numerical methods are used to analyze this relationship. A first step is to "normalize" the data, transforming each observation from scales convenient for measurement to a common scale, on which addition and multiplication can meaningfully combine the different observations. For example, z-transformation rescales every variable to the same scale, standardized variation from an expected value, but ignores scale differences between measured objects. Here we develop the concepts and properties of *projective decomposition*, which applies the same normalization strategy to both rows and columns by separating the matrix into row- and column-scaling factors and a scale-normalized matrix. We show that different scalings of the same scale-normalized matrix form an equivalence class, and call the scale-normalized, canonical member of the class its *scale-invariant form*. An important property of the scale invariant form $W$ of a matrix $A$ is that it preserves all pairwise relative ratios: that is, for any pair of rows $i \neq j$ and pair of columns $s \neq t$, $w_{is}w_{jt}a_{it}a_{js} = a_{is}a_{jt}w_{it}w_{js}$. Projective decomposition therefore provides a means of normalizing the broad class of "ratio-scale data", in which relative ratios are of primary interest, onto a common scale without altering the ratios of interest, and simultaneously accounting for scale effects for both organizations of the matrix values. Both of these properties distinguish it from z-transformation, which applies to interval-scale data and treats rows and columns asymmetrically.

## Introduction

In scientific experiments a real world phenomenon must first be quantified, i.e., represented numerically, before mathematical tools can be used to analyze the phenomenon. Some representational choices, including the choice of units for each value as well as their organization into a matrix or other mathematical structure, can influence the data analysis in a "nuisance" manner, i.e., one unrelated to the phenomenon of interest. Here we consider data *normalization* as the process of redefining the representation of the data to control at least some of these nuisance influences. For example, in standard Principal Components Analysis (PCA, [Pearson 1901; Hotelling 1933]), each quantified aspect of the phenomenon (each "variable" organized into a matrix column, with each "sample" organized into a matrix row) is normalized by the *z*-transformation

$$z_{ij} = \left(a_{ij} - \mu_j\right)/\sigma_j = \left(a_{ij} - E\{a_{*j}\}\right)/\sqrt{Var\left(a_{*j}\right)}.$$

This formula's linear dependence on the observed data ensures that the location (the phenomenon corresponding to $z = 0$) and scale (the phenomenon corresponding to $z_1 - z_0 = 1$) of the resulting z-scores correspond to the expected value and an observed one standard deviation difference, respectively, regardless of the units in which the phenomenon was observed. In multidimensional data, z-transformation also yields a common quantity that can be compared across columns, which may represent incomparable types of observations (e.g., one column of volumes and another of pressures).

When an observation is made in relative units, which provide no special correspondence between the mathematical concept of zero and the phenomenon represented by a zero value, z-transformation establishes such a correspondence: a zero z-score represents no difference from the statistically expected phenomenon. However, many observations are made in absolute units, in which the phenomenon quantified as zero is already mathematically equivalent to zero in a contextually appropriate, but non-statistical sense; a well-known example is temperatures quantified in Kelvin, where a zero value represents a relevant physical phenomenon, the total lack of kinetic energy. For such observations, z-transformation replaces a contextually appropriate meaning for zero with a different, statistical meaning. This has a strong effect on ratios of observed values, which are altered in a nonlinear way by changing the location of zero. Z-transformation is therefore more appropriate for "interval-scale" data, in which values are evaluated relative to a threshold by subtraction, than for "ratio-scale" data, which are evaluated via division.

Of course, the logarithm of two ratio-scale values converts their ratio to an interval, so it may seem that z-transformation can be applied equally well to ratio-scale data as to interval-scale data by taking logarithms prior to z-transformation. The equivalence between interval-scale data and the logarithm of ratio-scale data is valid only for a single dimension, however, and methods applied to multidimensional data, which involve rotations or other inherently multidimensional geometric operations, generally do not preserve the equivalence. Consider just two dimensions, in which the relationship between a rotational combination of two logarithms, $cos(\theta) log(x) + sin(\theta) log(y)$, and the logarithm of a rotational combination of two quantities, $log(cos(\theta)x + sin(\theta)y)$, is non-linear. This relationship does not lend itself to simple rules for finding a threshold value for the former, "normalized" values, which is equivalent to a threshold for the latter, original values, even though equivalent thresholds can easily be translated for the corresponding simple scaling transformation in a single dimension. Thus, while z-transformation of a logarithm does provide normalization of each individual dimension in a data matrix, it is not clear that common multidimensional analytical methods, even those as simple as principal components analysis, produce results which can be readily translated back to their real-world context, and z-transformation of logarithms should not be considered an optimal normalization method for multidimensional, ratio-scale data.

Another issue for data normalization concerns its relationship to the organization of the information. From an algebraic perspective, an $m \times n$ real-valued matrix $A$ defines correspondences between $R^m$ and $R^n$ accessed by multiplication on the left ($y = Ax \in R^m$) or right ($z = A^T y \in R^n$). In addition to organizing data, a data matrix therefore defines a correspondence between weighted combinations of the columns (in $Ax$, $x \in R^n$ weights the columns of $A$) and weighted combinations of the rows, and mathematical analysis determines properties of this correspondence. This mathematical correspondence is in essence symmetric: transposition of the matrix merely exchanges the roles of $m$ and $n$, and shouldn't be a factor that determines the results of a data analysis. A normalization method relates the observed data matrix to a normalized matrix through a set of constraints satisfied by the normalized matrix; the properties of the normalized matrix are then interpretable in terms of the constraints and its relationship to the original data matrix. For z-transformation, the relationship between the data matrix and the normalized matrix is described by the means and standard deviations of each column. This is an asymmetric set of constraints, treating rows differently than columns, and creates an asymmetry in the normalized matrix that was not present in the observed matrix.

However, there are many experiments in which the formal distinction between "variables" (columns) and "samples" (rows) reflects neither the nature of the phenomenon of interest nor the goals of the analysis. In exploratory data analysis, the analytical goals are often to identify clusters of samples, clusters of variables, or combinations of the two (biclusters) which merit further analysis, as well as quantitative relationships among rows, among columns, or between rows and columns; these goals are not inherently asymmetric. Likewise, high-dimensional data may be subject to unavoidable nuisance influences specific to rows as well as to columns. For example, in gene expression experiments the abundances $A$ of genes $g$ are measured in samples $s$ at either the RNA or protein level, and these measurements are subject to nuisance influences both gene-specific (*e.g.* chemical differences between genes leading to gene-dependent measurement sensitivity) and sample-specific (*e.g.* variation in sample handling). Moreover, it is typically of interest to compare expression of a gene across samples (genes in columns, $A = \{a_{sg}\}$) as well as within a sample across genes (samples in columns,

$A^T = \{a_{gs}\}$ ). In some experimental designs, care is taken to avoid nuisance influences associated with samples, e.g. by matching their properties across groups or excluding problematic samples. However, Big Data may be acquired by virtue of its availability rather than according to a carefully controlled experimental plan, and may require as much attention to nuisance influences across rows as down columns. As useful as *z*-transformation is, it does not provide a means to address this need: applying the z-transform twice in one order (*e.g.,* down columns and then across rows) leads to different results than in the opposite order (across rows and then down columns), and in neither case does such a process in general render an analysis invariant to even linear nuisance influences on the rows and columns simultaneously.

Here we define an alternative normalization methodology which normalizes the scale of a data matrix in the same manner for columns and rows simultaneously by means of a matrix decomposition. Unlike z-transformation, this approach does not provide an interpretation of the data in terms of statistical concepts such as expected values and standardized residuals; it is a geometric normalization rather than a statistical one. However, many data analysis methods commonly applied to *z*-scores involve geometric transformations such as rotations, and maintain their geometric interpretation when applied to geometrically normalized data. By applying the same scale constraints to rows (traditionally envisaged as geometric points) and columns (geometric lines or axes), this methodology supports the duality between points and lines of projective geometry, in which many proofs remain valid when the notions of point and line are interchanged[Dowling 1917]. In particular, this form of normalization preserves relative ratios, making it highly appropriate for normalization of ratio-scale data. We emphasize that this method is a matrix decomposition primarily to highlight that the full set of values required to reconstruct the original data matrix, which for z-transformation includes the column means and standard deviations, and in our methodology includes both row- and column-scaling factors, is part of the observed data and should be considered in the subsequent data analysis.

## Results

The contributions in this paper are mathematical concepts, and we therefore present them here as our results. We illustrate these concepts with simple examples in the Discussion section.

Two $m \times n$ matrices $A$ and $B$ are called *equivalent* if there is an invertible $m \times m$ matrix $P$ and invertible $n \times n$ matrix $Q$ for which $PB = AQ$. Equivalent matrices represent the same linear transformation under different choices of coordinate systems, with $P$ providing the coordinate transformation in $R^m$ and $Q$ in $R^n$. Invertible coordinate transformations include dilations (scaling each row or column by a positive scaling factor), reflections (scaling a row or column by -1), rotations, and combinations thereof. Our focus is on scale transformations without reflection or rotation, which correspond to matrices with positive diagonal elements and all non-diagonal elements $0$. We therefore make the following definition.

**Definition 1**. Two $m \times n$ matrices $A$ and $B$ are **equivalent up to scale** if there are strictly positive vectors $p \in R^n$ and $q \in R^m$ so that $D_q B = A D_p$ (or equivalently $B = D_{1/q} A D_p$), where $D_x$ denotes the square diagonal matrix with diagonal elements $d_{ii} = x_i$ from the vector $x$.

**Lemma 1.** *Equivalence up to scale is an equivalence relation.*
<u>Proof</u>. An equivalence relation must be reflexive, symmetric, and transitive.
*Reflexive:* For $p = 1_n$ and $q = 1_m$, $D_q A = I_m A = A = A I_m = A D_p$.
*Symmetric:* Suppose $A$ and $B$ are equivalent up to scale (*i.e.* satisfy the above definition in that order); then there exist strictly positive vectors $p \in R^n$ and $q \in R^m$ so that $D_q B = A D_p$. Let $p' = 1/p$ and $q' = 1/q$; note that $p' \in R^n$ and $q' \in R^m$ are also strictly positive vectors, $D_{q'} D_q = I_m$, $D_p D_{p'} = I_n$, and
$$D_{q'} A = D_{q'} A \left( D_p D_{p'} \right) = D_{q'} (A D_p) D_{p'} = D_{q'} (D_q B) D_{p'} = \left( D_{q'} D_q \right) B D_{p'} = B D_{p'}.$$
Thus, $B$ and $A$ are equivalent up to scale (in either order).
*Transitive:* Suppose $A$ and $B$ are equivalent up to scale and $B$ and $C$ are equivalent up to scale; then $A$, $B$, and $C$ are all of the same dimensions $m \times n$ and there are strictly positive vectors $p', p'' \in R^n$, $q', q'' \in R^m$ so that $D_{q''} C = B D_{p''}$ and $D_{q'} B = A D_{p'}$. Let $p = p'p''$ and $q = q'q''$; then $p$ and $q$ are both strictly positive, and
$$D_q C = D_{q'} D_{q''} C = D_{q'} B D_{p''} = A D_{p'} D_{p''} = A D_p,$$
and therefore $A$ and $C$ are equivalent up to scale. ∎

In parallel to the description given above for (rank) equivalent matrices, the class of matrices equivalent to a given matrix $A$ up to scale represents the same linear transformation under only the *dilations* of the coordinate axes in $R^m$ and $R^n$, i.e., changes of scale without any change of orientation. As a whole, the equivalence class can be considered the definition of a *scale-invariant linear transformation.* We now define a canonical member $W$ of each such equivalence class, as well as the relationship between $A$ and $W$, by means of a matrix decomposition with constraints on root mean square (RMS) as a measure of scale.

**Definition 2.** *The **projective decomposition** of an $m \times n$ matrix $A$ is the equation*
$$A = \{a_{ij}\} = \sigma \{\alpha_i w_{ij} \beta_j\} = \sigma D_\alpha W D_\beta,$$
where the components on the right hand side are

$\sigma = \sqrt{\sum_{ij} a_{ij}^2 / mn} = |A|$, the root-mean-square of the elements of $A$,

$\alpha \in R^m$ are strictly positive row scaling factors,

$\beta \in R^n$ are strictly positive column scaling factors, and

$W$ is an $m \times n$ matrix equivalent up to scale with $A$ satisfying the scale constraints

$|W_{i*}| = \sqrt{\sum_j a_{ij}^2 / n} = 1$ for every row $i$ (every row has unit RMS), and

$$|W_{*j}| = \sqrt{\sum_i a_{ij}^2/m} = 1 \text{ for every column } j \text{ (every column has unit RMS).}$$

These scale constraints also determine the root-mean-square of $W$:

$$|W| = \sqrt{\sum_{ij} a_{ij}^2/mn} = \sqrt{\sum_{i=1}^{m}\left(\sum_{j=1}^{n} a_{ij}^2/n\right)/m} = \sqrt{\sum_{i=1}^{m} 1^2/m} = 1.$$

We refer to the scale-normalized matrix $W$ as the unit-scale, canonical member of the class of matrices equivalent to $A$ up to scale, the *scale-invariant matrix* equivalent to $A$, or more simply the *scale-invariant form* of $A$.

For the $m \times n$ matrix $1_{m \times n}$ with 1 as every entry, each row satisfies $|1_n| = 1$, each column satisfies $|1_m| = 1$, and at the matrix level, $|1_{m \times n}| = 1$. While these constraints do not cause the *expected value* of every entry in $W$ to equal $1$, we take these observations about $1_{m \times n}$ as inspiration for the following definition.

**Definition 3.** *For any $m \times n$ matrix $M$, the **expected scale** $\|m_{ij}\|$ of entry $m_{ij}$ is the square root of the product of the root-mean-squares of the containing row and column,*

$$\|m_{ij}\| = \sqrt{|M_{i*}||M_{*j}|}.$$

Root-mean-square is a dimensionless concept of scale applicable to scalars, vectors, and matrices; in one dimension it is equivalent to the standard concept of scale (absolute value). The expected scale of every entry in a scale invariant matrix $W$ is $1$, and in this sense, the constraints above that define a scale invariant matrix ensure the expected scale of each matrix entry is equal to the (actual) scale of each row, each column, and the entire matrix.

When $W$ is the scale-invariant form of $A$, reorienting the rows of $A$ as matrix columns yields $A^T = \left(\sigma\, D_\alpha\, W\, D_\beta\right)^T = \sigma\, D_\beta^T\, W^T\, D_\alpha^T$, and we see $W^T$ is also the scale-invariant form of $A^T$ under the same scaling constants. Thus, projective decomposition normalizes the scale of rows and columns in precisely the same manner, as desired. Furthermore, for each pair $i \neq j$ of rows and pair $s \neq t$ of columns the entries of $A$ and its scale-invariant form $W$ satisfy the equation $w_{is}w_{jt}a_{js}a_{it} = w_{js}w_{it}a_{is}a_{jt}$. This equation can be rearranged into equivalent within-column or within-row "relative ratios" in $A$ that are preserved in $W$, e.g. $(w_{is}/w_{it})/(w_{js}/w_{jt}) = (a_{is}/a_{it})/(a_{js}/a_{jt})$, but in the all-numerator form its meaning for zero entries is well-defined. Preservation of all such relative ratios is a strong reason to consider projective decomposition as a scale normalization method for ratio-scale data, in which comparison of ratios is a primary analytical goal.

We can also further characterize the other components of the projective decomposition. Because the scale of $A$ is $\sigma = |A| = |\sigma\, D_\alpha\, W\, D_\beta| = \sigma |D_\alpha\, W\, D_\beta|$ and $W$ has unit scale,

$\left|D_\alpha\, W\, D_\beta\right| = 1 = |W|$ and therefore application of these scaling factors to $W$ does not result in a change of scale at the matrix level. We can therefore consider the effects of $\alpha$ and $\beta$ to be *relative* scaling among the rows and among the columns of $W$, respectively. Without further constraint $\alpha$ and $\beta$ are only defined up to a constant, as it is easy to verify that $D_\alpha\, W\, D_\beta = D_{g\alpha}\, W\, D_{\beta/g}$ for any positive constant $g$. If unique values for $\alpha$ and $\beta$ are desired, we suggest choosing $g$ to cause the concatenation of $\alpha$ and $\beta$, a vector of length $m + n$, to have unit scale.

At https://github.com/PriceLab/ProjectiveDecomposition we provide a straightforward projective decomposition algorithm, implemented in R[R Core Team 2013]. This algorithm is a variant of the Sinkhorn-Knopp matrix balancing algorithm[Sinkhorn 1967] as adapted by Rothblum *et al.*[Rothblum 1994], and belongs to the class of iterative proportional scaling algorithms. Matrix balancing refers to improving the numerical stability of matrix algorithms, but iterative proportional scaling algorithms are also used for statistical purposes, such as fitting the parameters of a log-linear model[Deming 1940]. Projective decomposition should not be confused with table raking[Ireland 1968], a more direct use of extended Sinkhorn-Knopp algorithms to control the row sums and column sums of contingency tables, rather than scale (root-mean-square). The conditions under which iterative proportional scaling algorithms converge are known for square matrices, but not for rectangular matrices; as a rule of thumb, the algorithm will converge unless the matrix is "too sparse" (discussed below). Depending on the sparseness pattern, even as few as two non-zero entries per row and per column may be sufficient for convergence.

In its original form, the Sinkhorn-Knopp algorithm decomposes a large class of non-negative square matrices into row- and column-scaling factors and a doubly-stochastic matrix, which is a matrix whose rows and columns each sum to $1$. This basic algorithm was rapidly extended to produce a specified set of row- and column-sums (under the necessary constraint that the sum of the row sums must equal the sum of the column sums) and to rectangular matrices[Menon 1968]. Note that by constraining sums of absolute values (i.e., the $L_1$-norm) rather than sums, the requirement that the matrix be non-negative can be eliminated. Rothblum *et al.* showed that by first taking the $p$-th power of each element (called the Hadamard $p$-th power, $A^{\circ p}$) of the non-negative matrix $A$, extended Sinkhorn-Knopp algorithms can rescale a matrix to control the $L_p$-norm of each row and column. Our projective decomposition algorithm takes this approach, iteratively rescaling the matrix to constrain the root-mean-square of each row and column to $1$ by constraining the $L_2$-norm of each row to $\sqrt{n}$ and each column to $\sqrt{m}$. We have successfully applied more sophisticated versions of this algorithm to sparse matrices with millions of entries; for matrices with billions of entries, it should be noted that a similar matrix balancing algorithm more suited to cloud computing has been described[Knight 2014], and could be adapted in the same manner to perform projective decomposition in a distributed data computing environment.

For square matrices, the Sinkhorn-Knopp algorithm and its extensions succeed if and only if the matrix has total support[Knight 2008], meaning the following conditions are met: (1) at least one element is non-zero; and (2) every non-zero entry is on at least one non-zero "permuted diagonal", where a permuted diagonal is understood to mean a vector containing exactly one entry from every row and every column. A "permuted diagonal" is ill-defined for rectangular matrices, since the diagonal must be of length $m$ and of length $n$, but $m \neq n$. The total support property indirectly limits the number of non-zero entries, and thus the sparsity of the matrix.

## Discussion

Data normalization is the task of placing observations of incomparable character on a common scale, so that mathematical tools can be used in a meaningful way for their analysis. Here we have described projective decomposition, a new, geometrically-motivated approach to normalization. In this section we hope to provide some intuition for why projective decomposition is suitable for normalization of ratio-scale data through the following simple, 2-dimensional examples; we trust it will not be difficult for the reader to extend this intuition from two dimensions to higher-dimensional examples. Since these examples contain a large number of points in just two dimensions, transposition of this data results in just two points in a large number of dimensions. the examples that follow are therefore not well-constructed to illustrate the important point that, on data consisting of a large number of points (rows) in a large number of dimensions (columns), projective decomposition satisfies the same geometric constraints on the rows and the columns.

We begin with an interval-scale example, on which z-transformation performs well (Figure 1). The data (Fig. 1, upper left) is a set of randomly generated points on small circles centered on a rectilinear grid of points (7x7 grid, 60 points per circle; 2,940 x 2 data matrix). We applied three normalization schemes (remaining plots), which show that z-scoring re-scales, but preserves relative intervals between all points regardless of direction along which the interval is measured, while first log-transforming the data (lower right) destroys this property. Projective decomposition (lower left) projects all data to the same distance from the origin, also destroying relative intervals.

We also examined this data in polar coordinates, plotted as independent angle and radius axes, before and after normalization (Figure 2). Ratios represent the slopes of lines through the origin, corresponding to angle (the x-coordinate) in these plots. Z-transformation, whether preceded by log-transformation (lower right) or not (upper right), preserves neither angles nor relative radius in this data. However, for this square data grid, the row scaling factors determined by Projective Decomposition (y-coordinate, lower left) are proportional to radius or RMS (y-coordinate, other three plots), and the angle of each point from the origin is precisely preserved (compare x-coordinate, lower left to upper left).

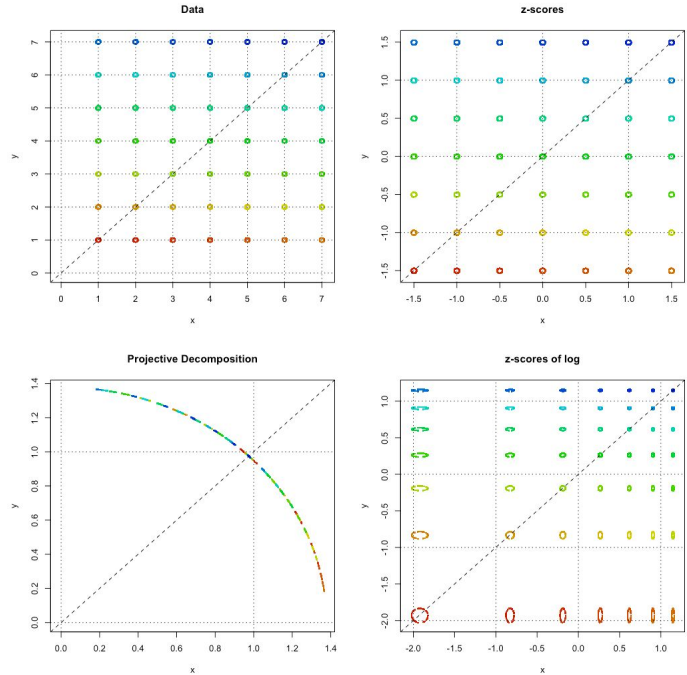

**Figure 1.** *Interval-scale data*: circles of points surrounding each intersection of a rectangular grid (upper left) and the same data normalized by z-transformation (upper right), projective decomposition (lower left), and log-transformation followed by z-transformation (lower right).

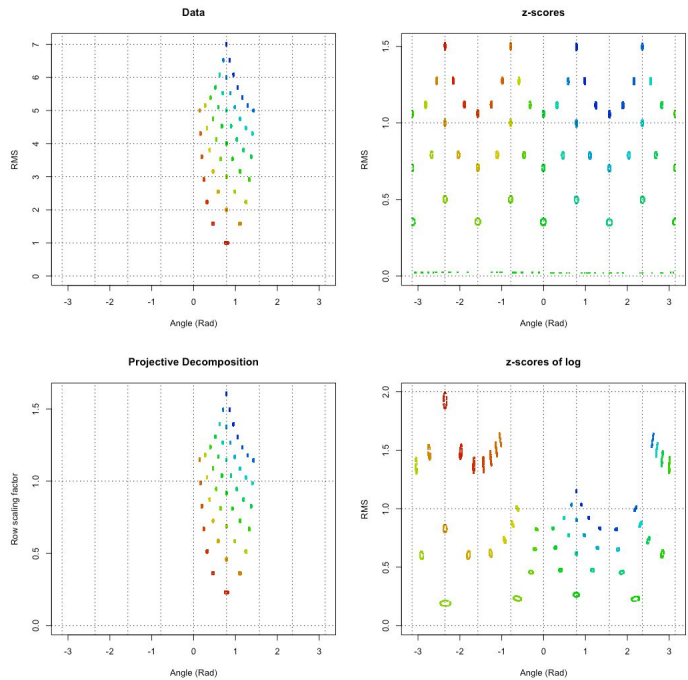

**Figure 2.** The data from Figure 1 shown in polar form. Each point has a row-scaling factor computed by projective decomposition; these are used as the RMS values in the lower left plot. However, angles are not in general precisely preserved by Projective Decomposition. Note that in our ratio-scale dataset (Figure 3), an x-y plot of the data (upper left) shows the x- and y-extents of the data differ by a 3-1 ratio. Z-transformation and projective decomposition both equalize this difference in scale between matrix columns corresponding to the x and y coordinates, as revealed by their common feature of x-y symmetry (Figure 3, upper right and lower left plots), unlike z-transformation after taking logarithms (Figure 3, lower right).

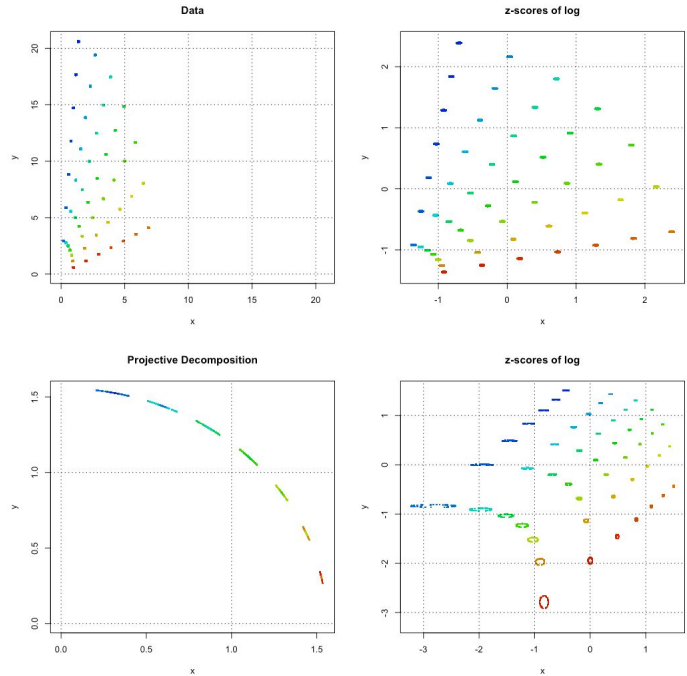

**Figure 3.** *Ratio-scale data*: circles of points surrounding each intersection of a radial grid (upper left) and the same data normalized by z-transformation (upper right), projective decomposition (lower left), and log-transformation followed by z-transformation (lower right).

Polar coordinate plots of the data from Figure 3 before and after normalization (Figure 4) reveals significant differences in the treatment of angles (ratios) by the three normalizations. Regardless of dataset, z-transformation preserves intervals up to scale (Figures 1 and 3, upper right), but neither angle nor radius (Figures 2 and 4, upper right). Preceding z-transformation with log-transformation, a common approach to normalization of ratio-scale data, places points the same direction from the origin on parallel lines with unit slope (Figure 3, lower right), however this does not result in preservation of either angle or radius (Figure 4, lower right). Furthermore, rotation of the data prior to normalization, corresponding to a simple horizontal shift on a polar plot, would have a very different effect prior to normalization than after this normalization (Figure 4, compare upper left and lower right). In contrast, projective decomposition results in preservation of the radial grid in this radially gridded data, and rotation of the data prior to normalization (lateral shifts of Figure 4, upper left) have an effect equivalent to rotation of the normalized data (lateral shifts of Figure 4, lower left).

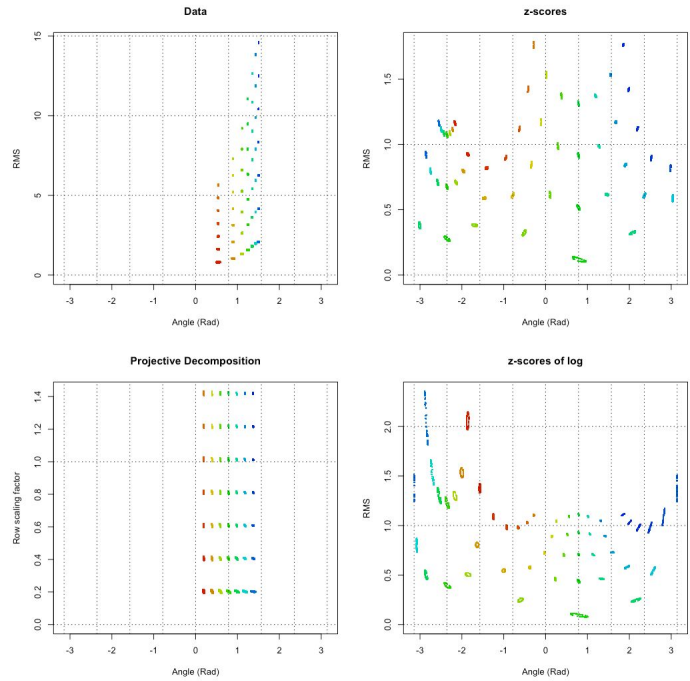

**Figure 4.** The data from Figure 3 shown in polar form. Each point has a row-scaling factor computed by projective decomposition; these are used as the RMS values in the lower left plot.

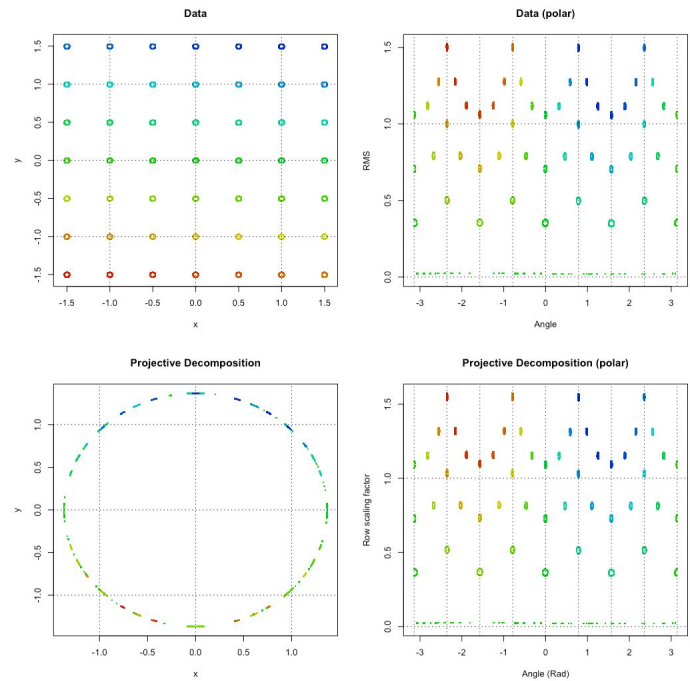

**Figure 5.** The z-scored interval data (upper left; see Figure 1, upper right) was renormalized by projective decomposition (lower left). Polar plots (data, upper right; normalized, lower right) demonstrate that, unlike log-transformation, projective decomposition is equally effective on data with mixed signs as on non-negative data.

A significant limitation of log-transformation is that the logarithm is restricted to strictly positive input values. In our final example (Figure 5), we have used the z-transformed version of our interval-scale data as a mixed-sign example. This data cannot be log-transformed, and (re-applying) z-transformation has no effect. Projective decomposition (lower left and right) is compatible with mixed-sign data, with the same preservation of angle and radius as for non-negative data in all four quadrants.

Projective decomposition has appropriate properties to be the normalization method of choice for ratio-scale data, and a viable alternative for simultaneous normalization of rows and columns for both interval-scale and ratio-scale data.